\documentclass[onecolumn,manuscript]{revtex4}
\usepackage{amssymb}

\usepackage{epsfig}
\usepackage[english]{babel}
\usepackage{latexsym}
\usepackage{graphics}
\usepackage{subfigure}
\usepackage{epsfig}
\usepackage{graphicx}
\usepackage{dcolumn}
\usepackage{amsmath}

\newcommand{\be}{\begin{equation}}
\newcommand{\ee}{\end{equation}}
\newcommand{\bey}{\begin{eqnarray}}
\newcommand{\eey}{\end{eqnarray}}
\newcommand{\bw}{\begin{widetext}}
\newcommand{\ew}{\end{widetext}}

\begin{document}                % INITIALIZE - DONT CHANGE % %

\title{Laser-assisted collision effect on nonsequential double ionization of helium in a few-cycle laser pulse}
%\rm
%\vskip0.8truecm

\author{Hongyun Li$^{1,2}$, J. Chen$^3$,  Hongbing Jiang$^{2}$,
Jie Liu$^3$, Panming Fu$^1$, Qihuang Gong$^{2}$, Zong-Chao
Yan$^{4, 5}$ and Bingbing Wang$^{1*}$}

\address{$^1$Laboratory of Optical Physics, Beijing National Laboratory for Condensed Matter Physics,
Institute of Physics, Chinese Academy of Sciences, Beijing 100080,
China}

\address{$^2$Department of Physics, State Key Laboratory for
Artificial Microstructures and Mesoscopic Physics, Peking
University, Beijing 100871, China}

\address{$^3$Center for Nonlinear Studies, Institute of Applied Physics and
Computational Mathematics, Beijing 100088, China}

\address{$^4$Department of Physics, University of New Brunswick, P. O. Box
4400, Fredericton, New Brunswick, Canada E3B 5A3}
\address{$^5$Center for Theoretical Atomic and Molecular Physics,
the Academy of Fundamental and Interdisciplinary Sciences,
Harbin Institute of Technology, Harbin 150080, China}

\begin{abstract}
Nonsequential double ionization (NSDI) of helium in an intense
few-cycle laser pulse is investigated by applying the
three-dimensional semi-classical re-scattering method. It is found
that the momentum distribution of He$^{2+}$ shows a
single-double-single peak structure as the pulse intensity
increases. According to the different mechanisms dominating the
NSDI process, the laser intensity can be classified into three
regimes where the momentum distribution of He$^{2+}$ exhibits
different characteristics. In the relatively high intensity
regime, an NSDI mechanism named the ``laser-assisted collision
ionization" is found to be dominating the NSDI process and causing
the single peak structure. This result can shed light on the study
of non-sequential ionization of a highly charged ion in a
relatively intense laser pulse.

\end{abstract}

\par\noindent

\maketitle

\par\noindent
$^*$ Corresponding author: wbb@aphy.iphy.ac.cn

\section{Introduction}

Study of non-sequential double ionization (NSDI) of an atom in a
laser field has become an important research topic in strong-field
physics~\cite{1,2,3,4} since its first experimental observation in
1983~\cite{5}. An NSDI process in an external laser field provides
us with an effective way to investigate and eventually control the
electron-electron correlations in a multi-electron atom or
molecule experimentally~\cite{1,2,3} or
theoretically~\cite{4,6,7,8,9}. At present, the widely accepted
theory of NSDI is the three-step re-collision
model~\cite{10,11,12}, in which one electron first escapes from
the atom by tunneling through the atomic-ground-state barrier,
formed by the Coulomb potential and the laser field. It is then
driven back by the laser field and collides with its parent ion,
resulting in ionization or excitation of a second electron.
According to this model, if the second electron is ionized
directly by a collision with the returning electron, the process
is called the collision-ionization (CI). Whereas if the second
electron is excited by a collision followed by an ionization in
the laser field through tunneling, the process is called the
collision-excitation-ionization (CEI). It is now clear that when
the laser intensity is so weak that the maximum returning kinetic
energy $E^{\rm max}_{\rm kin}$ is smaller than the ionization
potential of the second electron $I_p$~\cite{13}, the NSDI process
is dominated by the CEI mechanism, and the momentum distribution
of the double charged ion along the polarization of the laser
field shows a single peak at around the zero momentum. On the
other hand, when the intensity increases so that $E^{\rm max}_{\rm
kin}$ is larger than $I_p$, the NSDI is dominated by the CI
mechanism, and the momentum distribution shows a double-peak
structure with a minimum at the zero momentum. The distance
between the two peaks increases as the intensity increases, and
the extreme position of the right (left) peak is predicted to be
$4\sqrt{U_p}$ ($-4\sqrt{U_p}$ )~\cite{14}, where $U_p$ is the
ponderomotive energy of the electron in the laser field. However,
as the laser intensity further increases, whether or not the
positions of the two peaks can remain is still an open question.
Using the classical rescattering model, Feuerstein {\it et
al.}~\cite{14} predicted that these two peaks will remain at the
extreme values $\pm4\sqrt{U_p}$ for double ionization and the
momentum distribution of the multiple ionization will exhibit a
broad distribution with multiple peaks at $\pm2n\sqrt{U_p}$, where
$n$ is the number of the ionized electrons. However, the recent
experimental observation by Rudenko {\it et al.}~\cite{15} proved
that the two peaks shifts towards considerably lower values as the
intensity increases, and the distribution becomes a broader peak
before the sequential ionization occurs. Obviously, these
experimental results can not be simply attributed to the
well-known CI mechanism.

In this work, we employ three-dimensional semi-classical
re-scattering method~\cite{8,9} to investigate the intensity
dependence of the NSDI of helium in a few-cycle laser pulse. The
momentum distribution exhibits a single-double-single peak
structure as the intensity increases. In particular, a single peak
distribution at a weak laser intensity becomes a double-peak
structure and the positions of the two peaks increase to $\pm 4
\sqrt{U_p}$ as the intensity increases. When the intensity
increases further, the positions of the double peaks shift towards
zero from $\pm 4 \sqrt{U_p}$ and the two peaks finally become a
single peak. These results agree qualitatively well with the
experimental observations by Weber {\it et al.}~\cite{17} and
Herrwerth {\it et al.}~\cite{18} (Fig.~4 in that paper) for doubly
charged argon ions, as well as the results by Rudenko {\it et al.}
for neon ions~\cite{15,16}. Furthermore,through tracing the NSDI
trajectories, we find that, when the intensity increases from
intermediate- to strong-field regime, the NSDI mechanism changes
from CI to a new mechanism which we call the
``laser-assisted-collision-ionization" (LACI). In the strong-field
regime, the trajectories with small energy transition between the
two electrons during collision play a dominant role in the NSDI
with the help of the laser field, leading to the single peak
distribution. Our results can shed light on the study of
non-sequential ionization of the highly charged ions in a
relatively strong laser field.

The paper is organized as follows. In Sec.~II, we will briefly
review the basic theory of the three-dimensional semi-classical
re-scattering method. In Sec.~III, we will study the NSDI in
different intensity regimes. In Sec.~V, we will present the
conclusions.

\section{formulation}

We study the NSDI of helium in a linearly polarized few-cycle
laser pulse. A more detailed description of the three-dimensional
semi-classical recollision method may be found in~\cite{8,9}. Here
we briefly outline the formalism. Atomic units are used
throughout, unless otherwise specified. The linearly polarized
pulse is given by $\mathbf{E}(t)=-d\mathbf{A}(t)/dt$, where the
vector potential is
\begin{eqnarray}
&&\mathbf{A}(t)=(0,0,A_{0}\cos^{2}(\omega t/T)\sin(\omega
t+\varphi_{0})),
 \label{eq:1}
\end{eqnarray}
here $\omega$ is the carrier frequency, $T$ is the pulse duration,
and the carrier-envelope phase (CEP) of the pulse is
$\varphi_{0}=0$ in the calculations. The first electron is set
free by the external field with the weight given by the tunneling
probability calculated using the Ammosov-Delone-Krainov (ADK)
theory~\cite{19}. The subsequent evolutions of the ionized
electron and the remaining bound electron, driven by the combined
Coulomb potential and the laser field, are described by the
classical Newtonian equation
\begin{eqnarray}
{d^2 \mathbf{r}_j \over d t^2} = -\mathbf{E}(t)-
 \mathbf{\nabla}_j (V^{(j)}_{ne} +V_{ee}),
 \label{eq:2}
\end{eqnarray}
where $j=1$ and 2 corresponding to the ionized and bound electron
respectively. The Coulomb potential between the nucleus and the
electron is $V^{(j)}_{ne}=-2/|\mathbf{r}_j|$, and the Coulomb
potential between the two electrons is
$V_{ee}=1/|\mathbf{r}_1-\mathbf{r}_2|$.

To evolve Eq.~(\ref{eq:2}), we need to know the initial positions
and velocities of the two electrons. The initial position of the
ionized electron is determined by an equation which includes an
effective potential~\cite{8,9}. The initial velocity of this
electron is set to be $v_x=v_{\rm per}\cos\theta$, $v_y=v_{\rm
per}\sin\theta$, and $v_z=0$. The initial conditions for the bound
electron are determined by assuming that the bound electron is in
the ground state of He$^+$ and its initial distribution obeys the
micro-canonical distribution~\cite{20}. The weight of each
classical trajectory is proportional to $W(t_0, v_{\rm
per})=w(t_0)\bar{w}(t_0,v_{\rm per})$, where $w(t_0)$ is the
tunneling ionization probability at time $t_0$ given by the ADK
theory and the quantum mechanical transverse velocity
distribution~\cite{8,9} of the ionized electron is
\begin{eqnarray}
\bar{w}(t_0, v_{\rm per})=\frac{(2|I_p|)^{1/2}}{ | \mathbf{E}(t_0)|
\pi}\exp (- v^2_{\rm per}(2|I_p|)^{1/2}/ | \mathbf{E}(t_0)|),
\label{eq:3}
\end{eqnarray}
with $I_p$ being the ionization threshold of Helium. This
distribution is adopted to simulate the quantum diffusion of the
ionized wave packet moving in the external field.

We choose a starting time $t_{0}$, which is uniformly distributed
along the whole pulse, and then evolve Eq.~(\ref{eq:2}) to the end
of the pulse. In order to find out whether both electrons are
ionized, we calculate the final energy of each electron after that
the laser pulse has been turned off. If both electrons have
positive energies, then the NSDI has occurred. The frequency of
the laser pulse is $\omega=0.0578$~a.u. (the wavelength
$\lambda=760$~nm) and the pulse duration is $T=13.5$~fs which
contains five optical cycles. Figure 1 presents the ionization
yield of He$^{2+}$ ion as a function of the laser intensity. The
well-known ``shoulder structure" shown in Fig.~1 indicates that
the laser intensity considered in this work is below the
sequential double ionization threshold of helium.

\section{intensity dependence of NSDI in a few-cycle pulse}

We now consider the intensity dependence of NSDI of a helium atom
in a few-cycle pulse. Figure~2 presents the momentum distribution
of He$^{2+}$ parallel to the polarization of the laser field for
the peak intensity $I=2\times 10^{14}$ W/cm$^2$ (a), $2.5\times
10^{14}$ W/cm$^2$ (b), $3.5\times 10^{14}$ W/cm$^2$ (c),
$5.0\times 10^{14}$ W/cm$^2$ (d), $7.0\times 10^{14}$ W/cm$^2$
(e), and $1\times 10^{15}$ W/cm$^2$ (f). The distribution shows a
single-double-single peak structure as the intensity increases,
hence we classify the laser intensity into three regimes,
according to the characteristics of the momentum distribution of
He$^{2+}$ shown in Fig.~2. In the weak-field regime (Fig.~2(a)),
there is a single peak centered at the +$|P_z|$ direction which is
very close to zero momentum. In the intermediate-field regime
(Figs.~2(b)-2(e)), a double-peak structure is formed and the
distance between the two peaks changes with the intensity. The
maximum (minimum) position of the right (left) peak is
$4\sqrt{U_p}$ ( $-4\sqrt{U_p}$) at $I=3.5\times 10^{14}$ W/cm$^2$.
Finally in the strong-field regime (Fig.~2(f)), the momentum
distribution presents a single peak.

According to the three-step recollision model, there are three
factors which can influence the NSDI rate: (1) the tunneling rate
of the first electron e$_1$, (2) the kinetic energy that e$_1$ can
carry when it returns, and (3) the influence of the laser field on
the recollision process when e$_1$ returns. The relative role of
these three factors in determining the NSDI yield can be
identified by analyzing the momentum distribution of the He$^{2+}$
ion. Furthermore, the crucial influencing factor on NSDI, as well
as the corresponding mechanism, differs in different intensity
regimes.

\subsection{Weak- and intermediate-field regime}

We first consider NSDI in the weak-field regime. It is well known
that, when the intensity of the laser is so weak that the maximum
kinetic energy of the returning electron is smaller than the
ionization potential of He$^+$, the CEI mechanism dominates the
NSDI process, where there is an obvious time delay between the
recollision of the first electron and the ionization of the second
electron. We now perform a time-momentum analysis of the NSDI
process~\cite{18} to illustrate the NSDI in such a weak laser
field. Figure~3(a) presents the NSDI momentum distribution as a
function of time when $I=2\times 10^{14}$ W/cm$^2$. The horizontal
and vertical axes correspond to the time when NSDI occurs and the
momentum of He$^{2+}$ parallel to the laser field. The dashed
curve is the corresponding laser field. One can see that there is
only one group of NSDI trajectories appearing at about 0.5 laser
cycle where is one extreme of the laser field, as shown in
Fig.~3(a). We then trace back these NSDI trajectories to the
recollision time of the first electron as shown in Fig.~3(b),
where the recollision time is defined as the moment that the
collision between the two electrons happens. The horizontal and
vertical axes correspond to the recollision time and the kinetic
energy $E_{\rm kin}$ of the returning electron. Figure~3(b) shows
that the maximum value of $E_{\rm kin}$, which occurs at about
0.25 laser cycle, is smaller than the ionization potential of
He$^+$. Comparing Figs.~3(a) with 3(b), one can find that the time
delay between the double ionization and recollision is about $1/4$
laser cycle. Additionally, Fig.~3(b) shows that only the electron
of large kinetic energy which returns at about 0.25 laser cycle
contributes to the NSDI. This indicates that, in the weak-field
regime, where the CEI dominates the NSDI process, the returning
kinetic energy $E_{\rm kin}$ plays a crucial role in the NSDI
yield.

We next consider the NSDI in the intermediate-field regime where
the momentum distribution of the He$^{2+}$ ion displays a
double-hump structure. From Fig.~2 one can see that this
double-hump structure is well preserved over a large range of the
laser intensity, which qualitatively agrees with the COLTRIMS neon
data of Ullrich group~\cite{21}. Specifically, the intensity
region is from $2.5\times 10^{14}$ W/cm$^2$ to $8\times 10^{14}$
W/cm$^2$ in our calculations.

Similar as Fig.~3(a), Figure~4(a) illustrates the momentum
distribution of the He$^{2+}$ ion as a function of the NSDI time
with $I=3.5\times 10^{14}$ W/cm$^2$. For comparison, the laser
field $E(t)$ is also presented (the dashed curve). Figure~4(a)
shows that there are four groups of NSDI trajectories where the
double ionization (DI) occurs at approximately 0.2, 0.5, 0.7, and
1.0 laser cycle. Tracing back to the recollision time of the first
electron for these trajectories, we find that there are two
bunches of recollision trajectories corresponding to the four
groups of the DI trajectories, which is labeled by BI and BII in
Fig.~4(b). Furthermore, we find that the first and second groups
of the NSDI trajectories in Fig.~4(a) come from BI and the third
and forth groups come from BII. The double ionization of the first
and third groups in Fig.~4(a) occurs at about the zero crossings
of the laser field (0.2 and 0.7 laser cycle, respectively), with
almost no time delay comparing with the corresponding recollision
time of e$_1$. Obviously, this NSDI process is CI process. In
contrast, the second and forth groups of NSDI trajectories occur
at about the maximum (minimum) of the laser cycles with a time
delay of about $1/4$ laser cycle between the DI and the
recollision, indicating that these two groups are CEI process.
Moreover, the first and third groups of NSDI trajectories dominate
the contribution to NSDI, leading to a double peak structure in
the momentum distribution of He$^{2+}$ (Fig.~2(c)).

In the intermediate-field regime as shown in Fig.~4, since the
maximum values of the returning kinetic energy of e$_1$ for both
bunches BI and BII are larger than the ionization potential of
e$_2$, the kinetic energy is no longer the most crucial factor to
influence the NSDI yield. In contrast, the tunneling rate of e$_1$
plays the most important role in influencing the NSDI~\cite{22}.
Consequently, the third group from BII, which experiences a larger
tunneling rate, contributes more to the NSDI (Fig.~4(a)) than the
first group from BI, resulting in that the left peak is higher
than the right one in the momentum distribution in Fig.~2(c).

\subsection{ Intense-field regime}

We then further increase the intensity and find that the two peaks
of the momentum distribution merge into one peak when $I=1 \times
10^{15}$ W/cm$^2$, as shown in Fig.~2 (f). We also present the
momentum distribution of He$^{2+}$ as a function of time for $I=1
\times 10^{15}$ W/cm$^2$ in Fig.~5 (a). The dashed curve is the
corresponding laser field. There are four groups of NSDI
trajectories presented in Fig.~5(a). We also trace back these NSDI
trajectories to the recollision time of e$_1$. The kinetic energy
distribution versus collision time is presented in Fig.~5(b).
There are mainly three bunches of recollision trajectories,
labeled as BI, BII, and BIII in Fig.~5(b). The other weak bunches,
labeled as DI, DII, and DIII in Fig.~5(b), are from the multiple
recollision and have neglectable contributions to NSDI. Comparing
Figs.~5(a) with 5(b), we find that there is almost no time delay
between the recollision of e$_1$ and the ionization of e$_2$ for
the first three groups of the NSDI trajectories, whereas there is
about 1/4 laser cycle delay time between the forth group and the
corresponding recollision bunch BIII.  Furthermore, each one of
the first three groups presents a wide distribution, while the
main parts of the first two groups occur at about the maximum
(minimum) of the laser cycles, as shown in Fig.~5 (a). Comparing
these two groups with the first and third groups in Fig.~4(a), we
find that these two groups can not be attributed to the usual CI
mechanism. We hence define this new mechanism as the
``laser-assisted collision ionization" (LACI). The difference
between a CI and a LACI trajectory can be clarified as follows.
For a CI trajectory, the first electron is ionized with a
relatively lower tunneling rate (comparing with a LACI
trajectory), then it returns with a higher kinetic energy and
collides with the second electron at about the zero-crossing of
the laser field, where the influence of the laser on the
recollsion can be ignored, and the recollision process can roughly
be treated as a field-free collision process~\cite{23}. In
contrast, for a LACI trajectory, the first electron is ionized
with a relatively higher tunneling rate, then it returns with a
relatively lower kinetic energy and finally collides with the
second electron at about one extreme of the laser field, where the
laser field strongly affects the collision process and assists the
first electron to free the second electron. Furthermore, for a CI
trajectory, because the second electron is ionized by collision at
about the zero-crossing of the laser field, the final momentum of
the He$^{2+}$ ion is at about $\pm4\sqrt{U_p}$~\cite{14}, as shown
in Fig.~2 (c); In contrast, for a LACI trajectory, because the
second electron is ionized by collision at the extreme of the
laser field, the final momentum of He$^{2+}$ ion is at around
zero~\cite{14}, as shown in Fig.~2 (f).

In order to demonstrate that the LACI mechanism plays a dominant
role in leading to the single peak structure in the momentum
distribution in the intense-field regime, we re-calculate the NSDI
at $I=1 \times 10^{15}$ W/cm$^2$ by turning off the laser field
during the collision between the returning e$_1$ and the ion for
each NSDI trajectory. We turn off the laser field when the
returning electron is within $r_1=$5 a.u. from its parent ion. In
this way, we suppress the influence of the laser field on the
recollision process when e$_1$ passes through the ion. In order to
keep a small difference of the returning kinetic energy of e$_1$
for the two cases, we choose $r_1=$5 a.u. as the signal to turn
off the laser. In fact, we have obtained the similar results when
$r_1=$10 a.u. Figure~2(f) presents the momentum distribution with
(the solid curve) and without (the dashed curve) the laser field
during the recollision. It clearly shows that the contributions
from the trajectories with small returning kinetic energy of e$_1$
are suppressed when the laser field is turned off during the
collision, resulting in that the double-peak structure survives in
such a high laser intensity. This result confirms that the LACI is
the dominant NSDI mechanism in the intense-field regime and causes
the single peak structure in the momentum distribution of
He$^{2+}$. For comparison, we also present the momentum
distribution without the laser field (the dashed curve) during the
recollision for the case of $I=7\times 10^{14}$ W/cm$^2$ in
Fig.~2(e) and $I=3.5\times 10^{14}$ W/cm$^2$ in Fig.~2(c). For
both cases, the double peak structure remains when the laser field
is turned off during the collision, although the NSDI yield
decreases a little. It indicates that the influence of the laser
field on the recollision process does not play a key role on the
NSDI in the intermediate laser intensity regime.

To investigate in detail how the laser field affects the NSDI
yield through influencing the recollsion process, we present two
typical LACI trajectories in Figs.~6 and 7 for the recollision
electron e$_1$ (a) and the bound electron e$_2$ (b). For
comparison, Figs.~6(c) and 6(d) respectively show the
corresponding trajectories of e$_1$ and e$_2$ without the laser
field during the collision between the returning electron and the
ion core. As shown in Fig.~6(c), one can see that without the help
of the laser field, although the bound electron e$_2$ is set free
by the collision, the returning electron e$_1$ is recaptured by
the ion core; this is because that its small returning kinetic
energy can not overcome the Coulomb force of the ion. Hence in
this case, the laser field accelerates e$_1$ passing through the
ion core to avoid the Coulomb recapture of the ion.

Figure~7 presents another kind of LACI trajectory for e$_1$ (a)
and e$_2$ (b). Figs.~7(c) and 7(d) respectively show the
corresponding trajectories of e$_1$ and e$_2$ without the laser
field during the collision. As shown in Fig.~7(d), if the laser
field is turned off, although e$_1$ passes through the ion core
successfully, e$_2$ is still bounded by the ion core after the
collision; this is because that the small energy transition can
not set it free from the Coulomb bound state. Comparing Figs. 7(b)
with 7(d), one can finds that the acceleration along the $z$-axis
by the laser field during the collision helps e$_2$ get rid of the
binding of the ion. This can also be understood as that the laser
field lowers down the Coulomb barrier of the ion along the
$z$-axis during the collision, resulting in freeing e$_2$ after
the collision.

These two kinds of LACI trajectories indicate that the influence
of the laser field on the collision is twofold: on the one hand,
the laser field accelerates e$_1$ during the collision to avoid
the Coulomb recapture of the ion core; on the other hand, the
laser field lowers down the Coulomb barrier of e$_2$ along the
$z$-axis to provide it a chance to get free through the collision.

The experimental observation by Rudenko {\it et al.}~\cite{15} can
be clearly explained by the LACI mechanism. As mentioned
in~\cite{15}, when the laser intensity is larger than $4\times
10^{15}$~W/cm$^2$, the main channel for the Ne$^{3+}$ ion is the
sequential ionization of the first two electrons e$_1$ and e$_2$
followed by the non-sequential ionization of the third electron
e$_3$ by recollision of e$_2$ with the Ne$^{2+}$ ion. In that
intense-field regime, the laser accelerating effect on e$_2$ and
the lower-barrier effect on e$_3$ during the recollision influence
the NSDI trajectory effectively, leading to that the returning
trajectory, which collides with the ion at the extreme of the
laser field, starts to provide more contributions and finally
makes dominant contribution to the NSDI as the laser intensity
increases. As a result, the double peaks of the Ne$^{3+}$ ion in
the momentum distribution shift towards each other and eventually
merge into one peak as the intensity increases from $4\times
10^{15}$~W/cm$^2$ to $7\times 10^{15}$~W/cm$^2$ as shown in Fig.~1
of Ref.~\cite{15}.

\section{conclusions}

In this paper, we have presented a systematic study of NSDI of
helium in an intense few-cycle laser pulse by using the
three-dimensional semi-classical re-scattering method. According
to the different mechanisms dominating the NSDI process, the laser
intensity can be classified into three regimes where the momentum
distribution of He$^{2+}$ exhibits different characteristics. In
particular, the momentum distribution of He$^{2+}$ shows a
single-double-single peak structure as the pulse intensity
increases. In the relatively high intensity regime, a new NSDI
mechanism which is named the ``laser-assisted collision
ionization" is found to dominate the NSDI process and cause the
single peak structure before the sequential ionization occurs.
This result can explain the recent experimental observation of
Rudenko {\it et al.}~\cite{15}.

\begin{acknowledgments}

This research was supported by the National Natural Science
Foundation of China under Grant Nos. 60778009, 10634020 and
10521002, 973 Research Projects No. 2006CB806000 and No.
2006CB806003. ZCY was supported by NSERC of Canada. BW thanks
Jiangbin Gong for helpful discussions. HL thanks Xiaojun Liu for
helpful discussions.
\end{acknowledgments}

\begin{figure}
\includegraphics[width= \columnwidth]{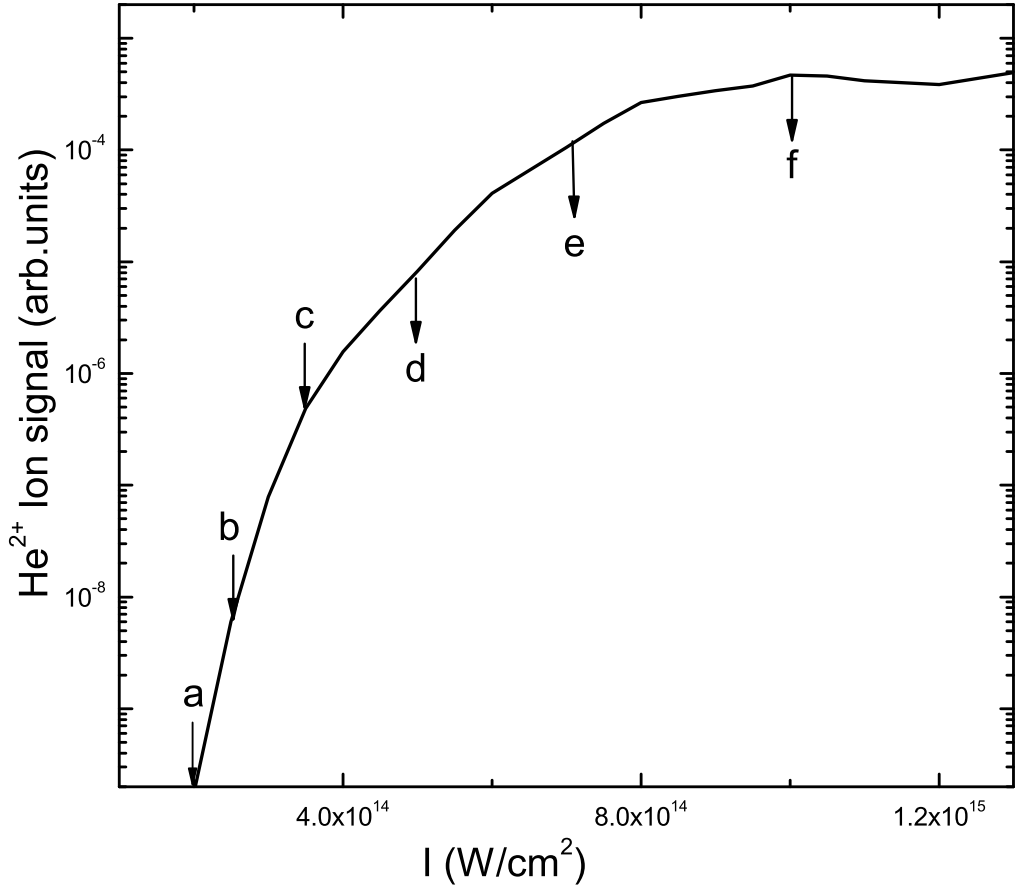} \vspace{-0.2cm}
\caption{Yield of He$^{2+}$ as a function of the peak
intensity of the laser pulse with $\varphi_{0}=0$.}
\end{figure}

\begin{figure}
\includegraphics[width= \columnwidth]{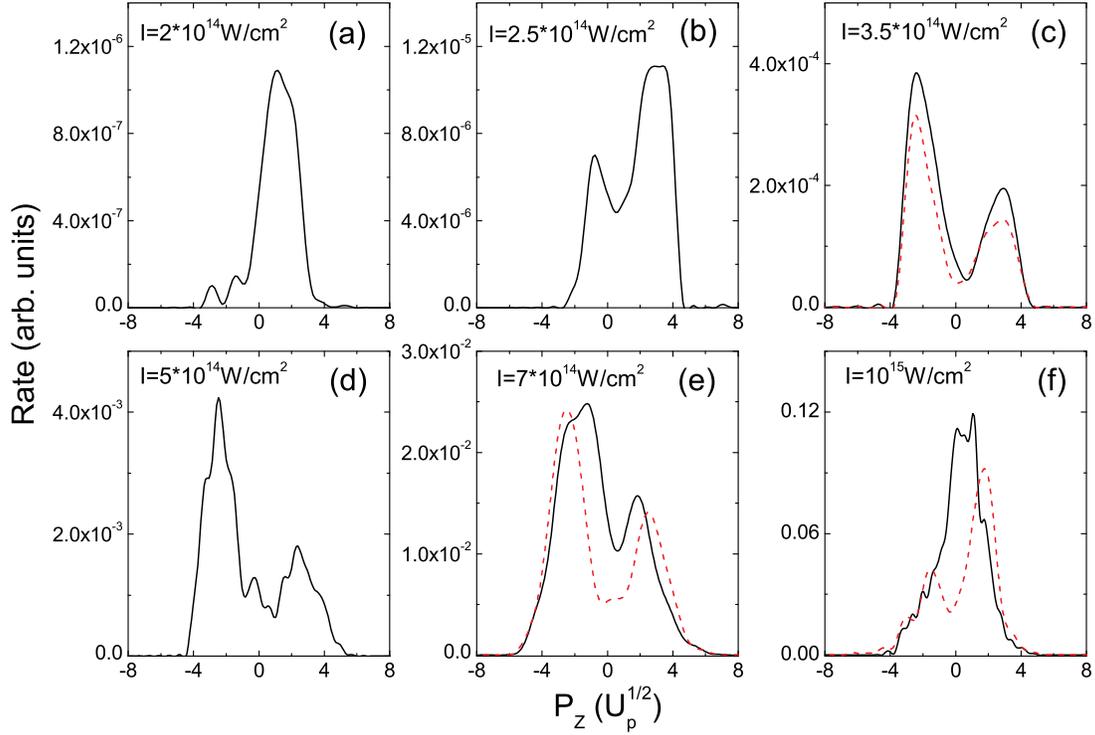} \vspace{-0.2cm}
\caption{(Color on line) Momentum distribution of He$^{2+}$
parallel to the polarization of the laser field with
$\varphi_{0}=0$ and $I=2.0\times10^{14}$~W/cm$^{2}$ (a),
$2.5\times10^{14}$~W/cm$^{2}$ (b), $3.5\times10^{14}$~W/cm$^{2}$
(c), $5\times10^{14}$~W/cm$^{2}$ (d), $7\times10^{14}$~W/cm$^{2}$
(e), and $1\times10^{15}$~W/cm$^{2}$ (f). The dashed curve in (c),
(e), and (f) is the momentum distribution of He$^{2+}$ without the
laser field during the recollision of e$_1$.}
\end{figure}
\begin{figure}
\includegraphics[width= \columnwidth]{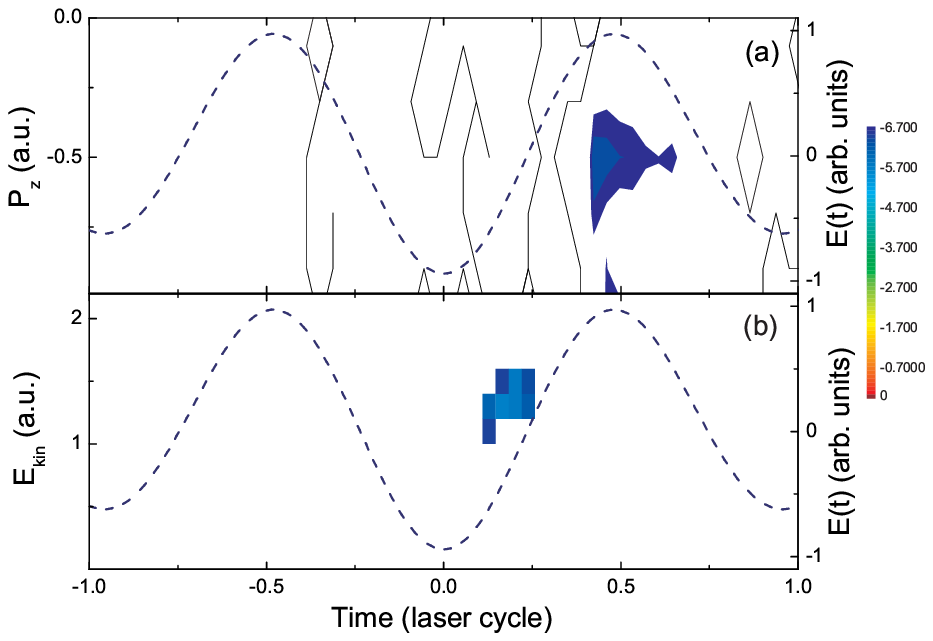} \vspace{-0.2cm}
\caption{(Color on line) (a) The momentum distribution of
He$^{2+}$ parallel to the polarization of the laser field as a
function of double ionization time with $\varphi_{0}=0$ when the
peak intensity of the laser pulse is
$I=2.0\times10^{14}$~W/cm$^{2}$. (b) The corresponding kinetic
energy distribution of the returning electron when it recollides
with the ion. The laser field is also presented (the dashed
curve). The results are plotted in log scale.}
\end{figure}
\begin{figure}
\includegraphics[width= \columnwidth]{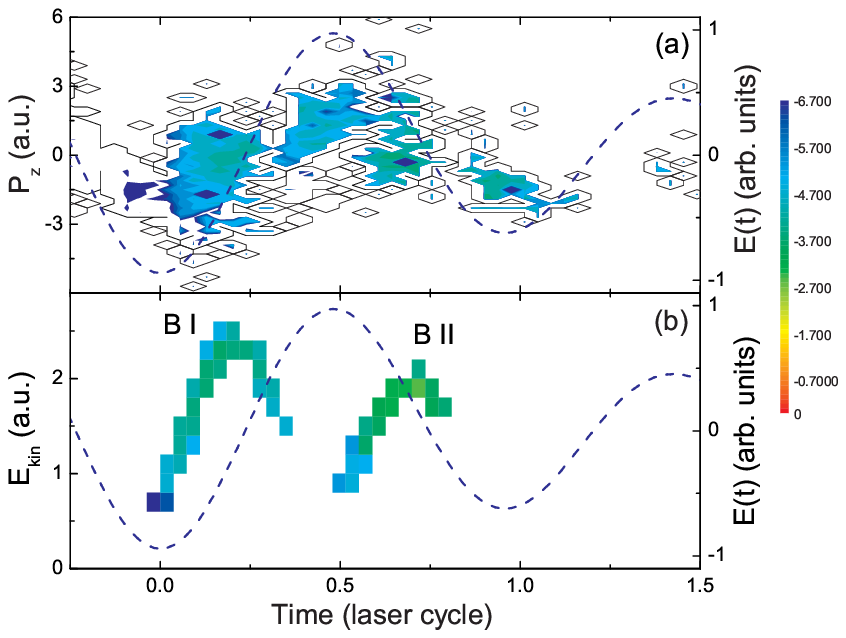}\vspace{-0.2cm}
\caption{(Color on line) (a) Momentum distribution of He$^{2+}$
parallel to the polarization of the laser field as a function of
double ionization time with $\varphi_{0}=0$ when the peak
intensity of the laser pulse is $I=3.5\times10^{14}$~W/cm$^{2}$.
(b) The corresponding kinetic energy distribution of the returning
electron when it recollides with the ion. The laser field is also
presented (the dashed curve). The results are plotted in log
scale.}
\end{figure}

\begin{figure}
\includegraphics[width= \columnwidth]{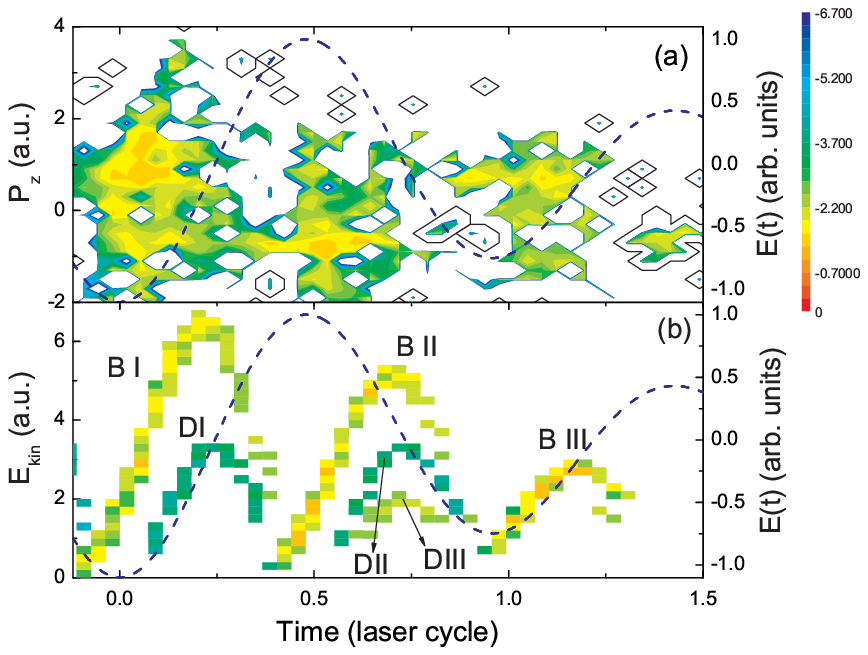} \vspace{-0.2cm}
\caption{(Color on line) (a) The momentum distribution of
He$^{2+}$ parallel to the polarization of the laser field as a
function of double ionization time with $\varphi_{0}=0$ when the
peak intensity of the laser pulse is $I=1\times
10^{15}$~W/cm$^{2}$. (b) The corresponding kinetic energy
distribution of the returning electron when it recollides with the
ion. The laser field is also presented (the dashed curve) in (a)
and (b). The results are plotted in log scale.}
\end{figure}

\begin{figure}
\includegraphics[width= \columnwidth]{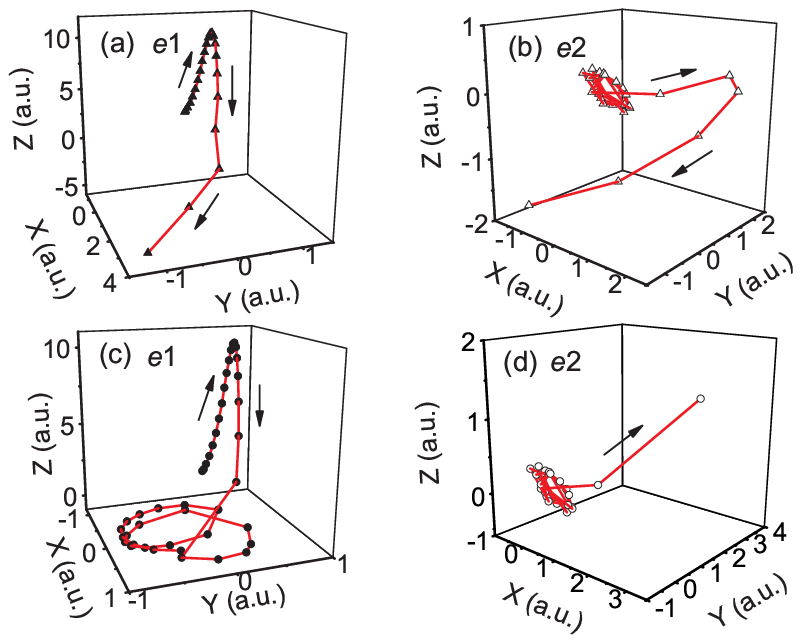} \vspace{-0.2cm}
\caption{(Color on line) An LACI trajectory for e$_1$ (a) and
e$_2$ (b) at $I=1\times10^{15}$~W/cm$^{2}$. (c) and (d): The
corresponding trajectories for e$_1$ and e$_2$ respectively
without the laser field during the recollision. The arrow
indicates the direction of a trajectory.}
\end{figure}

\begin{figure}
\includegraphics[width= \columnwidth]{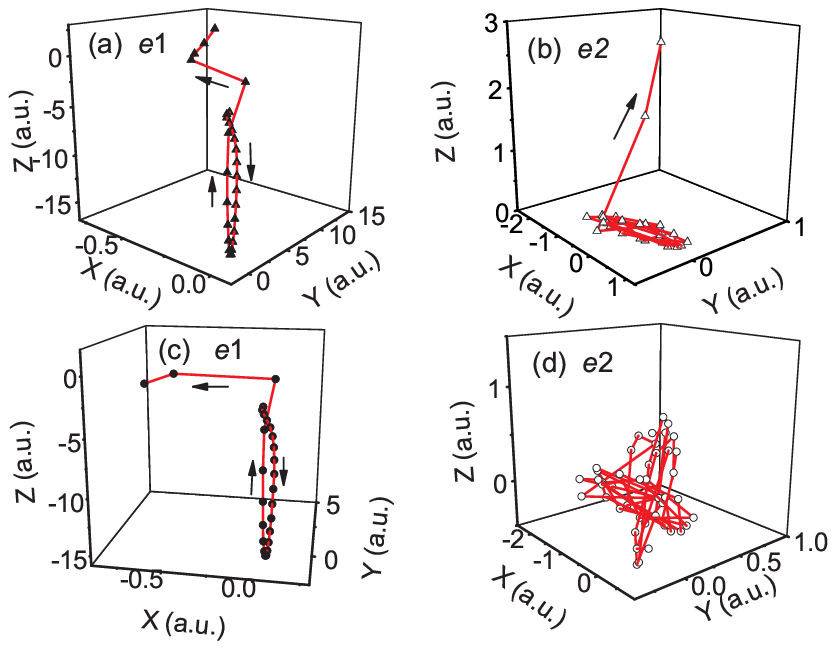} \vspace{-0.2cm}
\caption{(Color on line) Another LACI trajectory for e$_1$ (a) and
e$_2$ (b) at $I=1\times10^{15}$~W/cm$^{2}$. (c) and (d): The
corresponding trajectories for e$_1$ and e$_2$ respectively
without the laser field during the recollision. The arrow
indicates the direction of a trajectory.}
\end{figure}

\end{document}